\documentclass[aps,prd,preprint,superscriptaddress,nofootinbib]{revtex4-2}

\usepackage{graphicx}
\usepackage{silence}
\usepackage{orcidlink}
\usepackage{amsmath}
\usepackage{amssymb}

\providecommand{\apj}{Astrophys. J.}
\providecommand{\apjl}{Astrophys. J. Lett.}
\providecommand{\apjs}{Astrophys. J. Suppl. Ser.}
\providecommand{\aap}{Astron. Astrophys.}

\providecommand{\araa}{Annu. Rev. Astron. Astrophys.}
\providecommand{\mnras}{Mon. Not. R. Astron. Soc.}
\providecommand{\pasj}{Publ. Astron. Soc. Jpn.}
\providecommand{\pasp}{Publ. Astron. Soc. Pac.}
\providecommand{\nat}{Nature}

\begin{document}
\addtolength{\textheight}{3pt}

\title{Surrogate models for type II supernovae: Probing low-energy explosions and interaction-free regimes}



\author{Zhengyang Zhang \orcidlink{0009-0006-8370-3108}}
\email[Contact author: ]{zhangzhengyang@ynao.ac.cn}
\affiliation{International Centre of Supernovae (ICESUN), Yunnan Key Laboratory of Supernova Research, Yunnan Observatories, Chinese Academy of Sciences (CAS), Kunming 650216, People’s Republic of China}
\affiliation{University of the Chinese Academy of Sciences, 19A Yuquan Road, Shijingshan District, Beijing 100049, People’s Republic of China}

\author{Shuai Zha \orcidlink{0000-0001-6773-7830}}
\affiliation{International Centre of Supernovae (ICESUN), Yunnan Key Laboratory of Supernova Research, Yunnan Observatories, Chinese Academy of Sciences (CAS), Kunming 650216, People’s Republic of China}

\author{Nikhil Sarin \orcidlink{0000-0003-2700-1030}}
\affiliation{Kavli Institute for Cosmology, University of Cambridge, Madingley Road, CB3 0HA, United Kingdom}
\affiliation{Institute of Astronomy, University of Cambridge, Madingley Road, CB3 0HA, United Kingdom}

\author{Takashi J. Moriya \orcidlink{0000-0003-1169-1954}}
\affiliation{National Astronomical Observatory of Japan, National Institutes of Natural Sciences, 2-21-1 Osawa, Mitaka, Tokyo 181-8588, Japan}
\affiliation{Graduate Institute for Advanced Studies, SOKENDAI, 2-21-1 Osawa, Mitaka, Tokyo 181-8588, Japan}
\affiliation{School of Physics and Astronomy, Monash University, Clayton, Victoria 3800, Australia}

\author{Chengyuan Wu \orcidlink{0000-0002-2452-551X}}
\affiliation{International Centre of Supernovae (ICESUN), Yunnan Key Laboratory of Supernova Research, Yunnan Observatories, Chinese Academy of Sciences (CAS), Kunming 650216, People’s Republic of China}

\author{Bo Wang \orcidlink{0000-0002-3231-1167}}
\email[Contact author: ]{wangbo@ynao.ac.cn}
\affiliation{International Centre of Supernovae (ICESUN), Yunnan Key Laboratory of Supernova Research, Yunnan Observatories, Chinese Academy of Sciences (CAS), Kunming 650216, People’s Republic of China}


\received{3 February 2026}
\revised{16 June 2026}
\accepted{25 June 2026}

\begin{abstract}
To address the computational bottleneck of analyzing type II supernova samples from surveys like the Legacy Survey of Space and Time, we present two \textsc{stella}-based neural network surrogates: the interaction model for low-energy explosions with potential circumstellar material (CSM) interaction and the photospheric model for standard interaction-free SNe IIP. Each surrogate follows a two-stage design in which an autoencoder first compresses \textsc{stella} spectral energy distributions (SEDs) into a latent representation and a separate neural-network emulator then maps physical parameters to that latent space. Both models incorporate latent mixup regularization to improve latent-space continuity, while using different network backbones tailored to their respective regimes: ResNet blocks for the interaction model and 2D CNNs for the photospheric model. On test sets, the interaction model and photospheric model achieve normalized SED reconstruction MSEs of $\approx 9.1\times10^{-5}$ and $1.0\times10^{-4}$, respectively. For the low-luminosity SN~2005cs, the interaction model favors a low-mass progenitor ($M_{\mathrm{ZAMS}} = 10.40^{+0.04}_{-0.05}\,M_{\odot}$) with evidence for confined dense CSM, highlighting the likely presence of CSM interaction and offering a physical scenario consistent with direct imaging that helps resolve the historical mass discrepancy. For SN~2012aw, serving as a validation benchmark, the interaction model demonstrates consistency with previous studies by recovering a progenitor mass ($M_{\mathrm{ZAMS}} = 11.05^{+0.06}_{-0.06}\,M_{\odot}$). For the archetypal SN~1999em, the photospheric model derives a progenitor mass ($M_{\mathrm{ZAMS}} = 10.05^{+0.07}_{-0.04}\,M_{\odot}$) broadly consistent with direct preexplosion imaging limits without explicit CSM modeling, demonstrating that the photospheric model captures the essential physics of standard type IIP explosions. This model-level implementation can reduce full Bayesian parameter-inference runtime from days to minutes, providing a practical foundation for near-real-time physical characterization in large survey streams.
\end{abstract}


\maketitle

\section{Introduction}

Time-domain astronomy is entering an era of \emph{transient abundance}. Surveys such as the Zwicky Transient Facility (ZTF) and the Legacy Survey of Space and Time (LSST) will deliver orders-of-magnitude larger supernova samples over decade timescales, making \emph{fast, reproducible, and scalable} physics-based modeling and parameter inference a central requirement \citep{2019ApJ...873..111I,2019PASP..131a8002B,2020ApJ...904...35P}. Classical event-by-event Bayesian analyses using MCMC or nested sampling can become computationally prohibitive at survey scale, motivating alternative inference strategies, including simulation-based inference with normalizing flows and end-to-end software frameworks for electromagnetic transient inference \citep{2018ApJS..236....6G,2024MNRAS.531.1203S}.

Among core-collapse supernovae, hydrogen-rich type II supernovae (SNe II) are among the most common, yet they exhibit substantial diversity in light-curve morphology and spectral evolution \citep{2011MNRAS.412.1441L,2014MNRAS.442..844F,2014ApJ...786...67A,2015ApJ...799..208S,2016MNRAS.459.3939V,2017ApJ...850...89G,2019A&A...631A...8H,2022A&A...660A..40M,2026arXiv260203638E}. Observational studies have highlighted continuous variations across the classical SN~IIP/SN~IIL phenomenology in plateau shape, peak luminosity, and velocity evolution, while combined photometric and spectroscopic modeling emphasizes that similar broadband light curves can arise from different physical configurations, exacerbating parameter degeneracies when only limited-band photometry is available \citep{2013MNRAS.433.1745D,2019A&A...625A...9D,2019ApJ...879....3G,2020A&A...642A.143M,2022A&A...660A..41M}.

Interaction between SN ejecta and circumstellar material (CSM) strongly shapes early time observables and probes pre-SN mass loss. The underlying radiation-hydrodynamic framework—forward and reverse shocks converting kinetic energy into radiation—has long been established \citep{1982ApJ...258..790C,1994ApJ...420..268C,2017hsn..book..875C}. Although the most obvious signatures appear in narrow-line events (e.g., SNe~IIn/Ibn), interaction likely spans a broader SN~II population and may be linked to late-stage outbursts, winds, and binary evolution \citep{1990MNRAS.244..269S,2014ARA&A..52..487S,2014Natur.509..471G,2016ApJ...818....3K,2017hsn..book..403S,2024A&A...685A..58E}. In particular, confined dense CSM can delay shock breakout and reshape the earliest light-curve rise \citep{2011MNRAS.415..199M,2017A&A...605A..83D,2018NatAs...2..808F,2018MNRAS.476.2840M,2020MNRAS.496.1325B}.

Observational and modeling studies now suggest that such confined dense CSM is common among SNe~II and can systematically affect the first weeks after explosion \citep{2021ApJ...912...46B,2017ApJ...838...28M,2018ApJ...858...15M,2024ApJ...970..189J}. However, broadband light curves alone remain highly degenerate in CSM density structure and radial extent. Adding spectral information can break key degeneracies and better constrain CSM geometry and density profiles, while recent analytic scalings linking peak observables to physical parameters provide additional leverage on the dominant heating mechanism \citep{2019ApJ...878...56K,2023A&A...677A.105D}.

High-fidelity radiation-hydrodynamics codes such as \textsc{stella} \citep{1998ApJ...496..454B,2000ApJ...532.1132B,2006A&A...453..229B} can produce time-evolving SEDs, multiband light curves, and photospheric properties across a broad progenitor and explosion parameter space (e.g., ZAMS mass, explosion energy, $^{56}$Ni mass, mass-loss rate, and CSM extent/structure). However, they remain computationally expensive for large-sample inference. Our work builds on the \textsc{stella}-based SN~II model grids of \citet{2023PASJ...75..634M}, which provide the physical training foundation for fast surrogate modeling in the survey era. This grid has already supported survey-scale inference for ZTF and joint ZTF+ATLAS samples \citep{2023ApJ...945...46S,2024ApJ...969...57S,2025MNRAS.541..135H}. Related modeling efforts have also been applied to nearby well-observed events such as SN~2023ixf (with complementary radiative-transfer simulations) and to high-redshift SN~II samples discovered by JWST/JADES \citep{2024PASJ...76.1050M,2025ApJ...978...36F,2025ApJ...990..148H,2025A&A...694A.319K,2025RNAAS...9..135F,2025PASJ...77..851M,2025arXiv250105513C}.

While precomputed grids dramatically lower the barrier to physics-based modeling, accurate interpolation in continuous parameter space—especially for full time$\times$wavelength SEDs—remains a key bottleneck. Machine-learning surrogate models (emulators) have therefore emerged as a powerful route to accelerate expensive numerical predictions to millisecond-scale evaluations, including recent SN~II spectral-emulator work based on \textsc{tardis} \citep{2025A&A...702A..41V}. \citet{2025MNRAS.544.2653S} introduced surrogate models for SN~II light curves and photospheric properties trained on a large \textsc{stella} grid and discussed likelihood strategies that explicitly account for surrogate uncertainty in Bayesian fitting. However, extending accurate inference to the low-energy and low-mass regime remains a crucial challenge. Low-luminosity SNe~II make up a substantial fraction of the volumetric rate and provide a unique probe of the lower-mass limit of core-collapse progenitors, but they lie near parameter-space boundaries that are often sparsely sampled and numerically unstable \citep{2014MNRAS.439.2873S}.

In this work, we focus on the robust generation and inference of SN~II light curves and present two independent surrogate models: the interaction model, which is designed to provide stable coverage of low-energy and low-luminosity regimes with possible CSM interaction, and the photospheric model, which provides dedicated support for models without circumstellar interaction. Our central objective is to construct a surrogate capable of accurately generating time-evolving spectra and multiband light curves; these tools have been integrated into \textsc{redback}\footnote{\url{https://github.com/nikhil-sarin/redback}}, an open-source Bayesian inference software package for electromagnetic transients \citep{2024MNRAS.531.1203S}. Both models adopt a two-stage architecture. For the interaction model, we first train an autoencoder to compress SEDs into a 256-dimensional latent space, followed by a parameter-to-latent emulator. For the photospheric model, we employ a 2D convolutional autoencoder to compress high-dimensional SEDs into a low-dimensional latent representation, followed by a similar parameter-to-latent emulation step. To enhance continuity and robustness, both architectures introduce latent-space interpolation-consistency regularization (latent mixup). Furthermore, we incorporate time-weighted losses and gradient constraints to improve reconstruction fidelity, particularly during the critical early time and transition phases.

The paper is organized as follows. Section \ref{sec2} details the methodology, including the treatment of the training dataset, the specific architectures of the interaction model and photospheric model surrogates, and the implementation of the latent mixup regularization. Section \ref{sec3} presents the validation results, quantifying the spectral reconstruction fidelity and demonstrating the Bayesian parameter inference on three benchmark supernovae (SN~2005cs, SN~2012aw, and SN~1999em). Finally, Sec.~\ref{sec4} summarizes our findings, discusses the computational advantages, and addresses current physical limitations.

\section{Method}\label{sec2}
\subsection{Model overview}
Figure~\ref{fig:architecture} summarizes the network architecture used in this work. Both surrogate models adopt a two-stage autoencoder--emulator design: the autoencoder first compresses each time--wavelength spectral energy distribution (SED) into a compact latent representation and learns a nonlinear decoder back to spectral space; meanwhile, an independent emulator maps the physical input parameters to this latent space so that the frozen decoder can generate a complete SED during inference. The interaction model is designed for regimes affected by CSM interaction, including low-mass and low-energy explosions, and uses ResNet-based encoder, decoder, and emulator networks to model the interacting SED manifold. The photospheric model targets the classical interaction-free type~IIP regime. It uses a two-dimensional convolutional neural network (CNN) as its backbone. In both cases, the autoencoder input is a two-dimensional synthetic SED image. Before training, the \textsc{stella} spectra are interpolated onto a fixed $100\times100$ time--wavelength grid, so each input sample contains 10,000 normalized values. The two models have the same input dimensionality, but their grids are adjusted according to the relevant physical mechanisms. For the interaction model, the time axis spans 0.1--400 days and the wavelength axis is geometrically spaced from 500 to 49,500\,\text{\AA}. For the photospheric model, the time axis is sampled from 0.1 to 200 days and the preprocessing uses a linearly spaced wavelength grid over the same 500--49,500\,\text{\AA} range. The physical parameter ranges and valid model subsets differ, as summarized in Table~\ref{tab:model_parameters}.

Our implementation differs from the setup of \citet{2025MNRAS.544.2653S} in three main respects. First, we use the autoencoder latent representation directly instead of adding a PCA compression step; this avoids imposing a linear projection on an intrinsically nonlinear SED manifold and allows the latent coordinates to be optimized for spectral reconstruction without an additional PCA-to-decoder mapping. Second, we regularize the learned manifold with latent-space mixup, motivated by the broader manifold-mixup approach for enforcing smoother interpolations in learned representations \citep{2018arXiv180605236V}; the ablation test for this choice is presented in the Appendix, consistent with previous work showing that interpolation regularization can improve autoencoder latent spaces \citep{2018arXiv180707543B}. Third, we extend the surrogate coverage toward lower-mass and lower-energy explosions and add the photospheric model as a dedicated interaction-free surrogate for standard type~IIP light curves.

The interaction model adopts a confined CSM density structure characterized by a wind acceleration law \citep{2023PASJ...75..634M}. The CSM density $\rho_{\mathrm{CSM}}(r)$ is defined as

\begin{equation}
    \mathbf{\rho_{\mathrm{CSM}}(r) = \frac{\dot{M}}{4\pi r^2 v_{\mathrm{wind}}(r)},}
\end{equation}

\noindent where $\dot{M}$ is the mass-loss rate. The velocity profile $v_{\mathrm{wind}}(r)$ follows a $\beta$-law acceleration

\begin{equation}
    \mathbf{v_{\mathrm{wind}}(r) = v_0 + (v_\infty - v_0) \left( 1 - \frac{R_0}{r} \right)^\beta.}
\end{equation}

\noindent Here, $R_0$ is the progenitor radius, $v_0$ is the initial velocity at the surface, and $v_\infty$ is the terminal wind velocity (fixed at $10\,\mathrm{km\,s}^{-1}$). The parameter $\beta$ determines the efficiency of wind acceleration; a higher $\beta$ corresponds to a slower acceleration, resulting in a steeper density gradient in the immediate vicinity of the progenitor.
\begin{table*}[ht!]
    \centering
    \caption{Input physical parameters and their valid ranges for the interaction model and photospheric model surrogates. The last column summarizes how the interaction model differs from \citet[][Table~1]{2025MNRAS.544.2653S}. Note that the photospheric model includes derived parameters ($M_{\mathrm{env}}$, $R_0$) dependent on $M_{\mathrm{ZAMS}}$ (marked with superscript $a$), while the interaction model includes specific CSM interaction parameters. Mixing definitions: ``cm'' (central mixing) denotes no $^{56}$Ni mixing into the hydrogen-rich envelope; ``fm'' (full mixing) indicates a fully mixed $^{56}$Ni distribution throughout the entire ejecta; ``hm'' (half mixing) represents an intermediate extent. The interaction model inherently adopts the ``hm'' configuration.}
    \label{tab:model_parameters}
    \resizebox{\textwidth}{!}{%
    \begin{tabular}{l l c c c l}
    \hline
    \hline
    Parameter & Symbol & Unit & Interaction model range & Photospheric model range & Interaction model vs. Sarin+25 \\
    \hline
    \multicolumn{6}{c}{\textit{Progenitor \& Explosion Parameters}} \\
    \hline
    Progenitor Mass & $M_{\mathrm{ZAMS}}$ & $M_{\odot}$ & $9 - 18$ & $10 - 18$ & Lower bound extended (10 $\rightarrow$ 9) \\
    $^{56}$Ni Mass & $M_{\mathrm{Ni}}$ & $M_{\odot}$ & $0.001 - 0.3$ & $0.0001 - 0.3$ & Unchanged \\
    Explosion Energy & $E_{\mathrm{SN}}$ & $10^{51}$ erg & $0.1 - 5.0$ & $0.2 - 5.0$ & Lower bound extended (0.5 $\rightarrow$ 0.1) \\
    Mixing Parameter & \texttt{mixing} & -- & \textit{Fixed (hm)} & \{cm, fm, hm\} & Explicitly fixed to hm \\
    \hline
    \multicolumn{6}{c}{\textit{CSM configuration (interaction model only)}} \\
    \hline
    Mass-loss Rate & $\log_{10}(\dot{M})$ & $\log(M_{\odot}\,\mathrm{yr}^{-1})$ & $-5 \sim -1$ & -- & Unchanged \\
    CSM Radius & $R_{\mathrm{CSM}}$ & $10^{14}$ cm & $1 - 10$ & -- & Unchanged \\
    Density Slope & $\beta$ & -- & $0.5 - 5.0$ & -- & Unchanged \\
    \hline
    \multicolumn{6}{c}{\textit{Derived Envelope Parameters (photospheric model only)}} \\
    \hline
    Envelope Mass & $M_{\mathrm{env}}$ & $M_{\odot}$ & -- & $7.175 - 9.5$\footnotemark[1] & N/A \\
    Progenitor Radius & $R_0$ & $R_{\odot}$ & -- & $414 - 970$\footnotemark[1] & N/A \\
    \hline
    \end{tabular}%
    }
    \footnotetext[1]{These are derived photospheric-model parameters that depend on $M_{\mathrm{ZAMS}}$.}
\end{table*}

\subsection{Interaction model}
\subsubsection{Data preprocessing and augmentation}
The interaction model dataset is based on the \textsc{stella} radiation-hydrodynamics simulation grid of \citet{2023PASJ...75..634M}, supplemented by low-energy and low-mass models obtained from Moriya \citep{MoriyaPrivate}. After filtering invalid or incomplete outputs, the Interaction dataset contains 298,375 original valid SED samples on a uniform $100 \times 100$ time-wavelength grid. Each sample is defined by six physical parameters, including $M_{\mathrm{ZAMS}}$, $M_{\mathrm{Ni}}$, $E_{\mathrm{SN}}$, $\log_{10}\dot{M}$, $R_{\mathrm{CSM}}$, and $\beta$, together with the corresponding time-dependent \textsc{stella} SED. The spectral data were interpolated onto the uniform $100 \times 100$ grid and log-normalized to the $[0,1]$ interval.

To enhance model robustness, we applied dimensionless Gaussian perturbations with $\sigma=0.01$ in the normalized training space. Specifically, noise was added both to the standardized input-parameter vector $X$ and to the output array $Y$ (the $\log_{10}L_\nu$ SED values normalized to the $[0,1]$ interval), generating five augmented copies for each original sample. Including the original samples and their augmented copies, this gives a total of 1,790,250 samples.

The models were split into training, validation, and test groups with an 80\%/10\%/10\% ratio. Each original model and all of its Gaussian-perturbed augmented copies were assigned exclusively to the same subset. This prevents augmented versions of the same physical model from crossing split boundaries and therefore avoids data leakage between the training, validation, and test sets.

\begin{figure*}[t] 
    \centering 
    \includegraphics[width=0.95\textwidth]{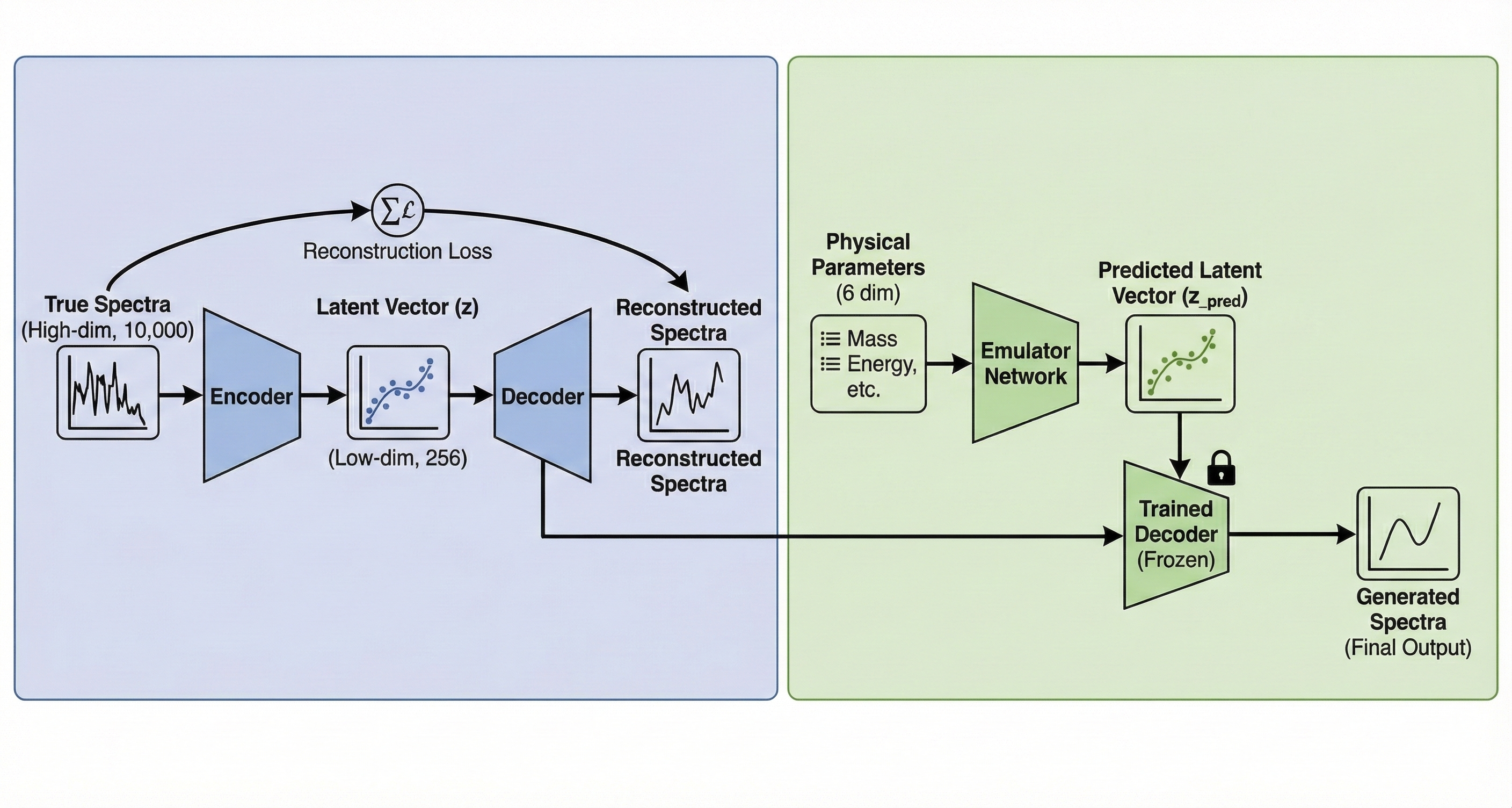}
    \caption{Schematic illustration of the two-stage surrogate modeling framework. This schematic is AI-generated for visualization purposes only. The two panels display the workflows for the spectral compression (Stage 1; blue background) and physical parameter emulation (Stage 2; green background) phases. The left panel shows the AutoEncoder architecture, which compresses high-dimensional true spectra into a low-dimensional latent vector ($\mathbf{z}$) by minimizing the reconstruction loss. The right panel shows the emulator training process, where the network maps physical parameters to a predicted latent vector ($\mathbf{z}_{\text{pred}}$). Note that the pretrained decoder is frozen (indicated by the padlock icon) during this stage to ensure consistent spectral generation from the predicted latent codes.
    \label{fig:architecture}}
\end{figure*}

\subsubsection{Spectral compression via latent mixup}
The primary objective of the first stage, as illustrated in the blue left panel of Figure~\ref{fig:architecture}, is to compress the high-dimensional spectral input $\mathbf{x} \in \mathbb{R}^{10000}$ into a low-dimensional latent vector $\mathbf{z} \in \mathbb{R}^{256}$. Both the encoder and decoder are constructed with 12 residual blocks, utilizing SiLU activation functions to capture nonlinear spectral features.

Standard autoencoders often suffer from discontinuous or fragmented latent spaces \citep{2018arXiv180707543B}. This irregularity makes it difficult for the model to accurately map continuous physical parameters to latent codes during the second stage. To resolve this mapping challenge, we introduce a regularization technique termed ``latent mixup consistency'' \citep{2018arXiv180605236V}. Its core concept is to ensure that linear interpolation in the latent space directly corresponds to a linear transition in the input spectral space. The total loss function for this stage is defined as

\begin{equation}
    \mathcal{L}_{\text{total}} = \mathcal{L}_{\text{recon}}(\mathbf{x}, \hat{\mathbf{x}}) + w_{\text{mixup}} \cdot \mathcal{L}_{\text{mixup}},
    \label{eq:stage1_total_loss}
\end{equation}

\noindent where $\mathcal{L}_{\text{recon}}$ enforces spectral fidelity to the original \textsc{stella} outputs, while $w_{\text{mixup}}$ sets how strongly the interpolation regularization is applied. In practice, this balances two goals: accurate spectrum reconstruction and a smooth latent manifold that is easier to map from physical parameters. The Mixup consistency term is defined as

\begin{equation}
    \mathcal{L}_{\text{mixup}} = \left\| \mathcal{D}(\lambda \mathbf{z}_1 + (1-\lambda)\mathbf{z}_2) - (\lambda \mathbf{x}_1 + (1-\lambda)\mathbf{x}_2) \right\|^2,
    \label{eq:mixup_loss}
\end{equation}

\noindent Here, $\mathcal{D}$ denotes the decoder; $\mathbf{z}_1$ and $\mathbf{z}_2$ are the latent codes of two samples; $\mathbf{x}_1$ and $\mathbf{x}_2$ are the corresponding input spectra; and $\lambda \sim \text{Beta}(\alpha, \alpha)$ is the random mixing coefficient \citep{2017arXiv171009412Z}. The parameter $\alpha$ controls the shape of the Beta distribution from which $\lambda$ is drawn: smaller values place the mixed sample closer to one of the two original samples, whereas larger values concentrate $\lambda$ closer to their average. In this work, we set $\alpha=0.2$. Physically, this term encourages intermediate points in latent space to decode into intermediate spectra, rather than producing nonphysical artifacts between neighboring models. This improves interpolation across sparsely sampled regions of parameter space and stabilizes the downstream inference of progenitor and explosion properties.

\subsubsection{Physical parameter emulation}
The second stage, shown in the green right panel of Figure~\ref{fig:architecture}, learns the mapping from the six-dimensional physical parameter space $\mathbf{p} \in \mathbb{R}^6$ to the latent code $\mathbf{z}$. For this purpose, we construct an emulator network consisting of 12 ResNet blocks with SiLU activations between layers.

During this phase, the pretrained decoder from Stage 1 remains frozen to preserve the learned manifold (as indicated by the padlock icon in Figure~\ref{fig:architecture}). The emulator predicts the latent code $\hat{\mathbf{z}}$, which is then passed through the frozen decoder to generate the reconstructed spectrum. The training objective minimizes a composite loss function consisting of a Huber loss in latent space and an L1 loss in spectral space.

\subsection{Photospheric model}\label{sec:clean_model}

The foundational dataset comprises 1,032 \textsc{stella} models (Moriya \citep{MoriyaPrivate}), each defined by six physical parameters ($M_{\mathrm{ZAMS}}$, $M_{\mathrm{Ni}}$, mixing parameter, $E_{\mathrm{exp}}$, $M_{\mathrm{env}}$, and $R_0$). To achieve robust training despite this relatively limited sample size, we implemented a 10$\times$ data augmentation strategy by injecting dimensionless Gaussian noise with $\sigma=0.01$ in the normalized training space, applied both to the standardized input-parameter vector $X$ and to the $[0,1]$-normalized $\log_{10}L_\nu$ SED output array $Y$, expanding the total dataset to 10,320 samples. The photospheric model dataset was split into training, validation, and test sets using an 80\%/10\%/10\% ratio, corresponding to 8,256, 1,032, and 1,032 augmented samples, respectively.

Regarding the network backbone, we observed that standard ResNet architectures yielded suboptimal performance on this specific dataset. Under the same setting (ResNet vs. CNN), ResNet shows a 42.15\% higher test-set MSE than CNN. Consequently, we adopted a 2D CNN. This architecture retains the established two-stage training strategy (as illustrated in Figure~\ref{fig:architecture}) and incorporates latent mixup regularization to ensure the smoothness and continuity of the latent manifold.
\section{Result}\label{sec3}

\subsection{Model performance}\label{sec:reconstruction}


To quantitatively assess the generative accuracy, we evaluated both stages of the surrogate pipeline on held-out test sets: the Stage~1 autoencoder, which measures the intrinsic SED compression--reconstruction accuracy, and the Stage~2 emulator, which measures the full parameter-to-SED surrogate accuracy after mapping physical parameters into the learned latent space. Errors are reported in the normalized training space ($Y$, the $[0,1]$-normalized $\log_{10}L_\nu$ array) and after conversion back to logarithmic luminosity units, expressed in dex.

For the interaction model, the Stage~1 autoencoder achieves $\mathrm{MSE}_Y=8.857866\times10^{-5}$, and the complete Stage~2 emulator gives a very similar $\mathrm{MSE}_Y=9.134600\times10^{-5}$, indicating that most of the error budget is already set by the autoencoder reconstruction limit rather than by the parameter-to-latent mapping. For the photospheric model CNN, the Stage~1 and Stage~2 test errors are likewise nearly identical, with $\mathrm{MSE}_Y=1.001196\times10^{-4}$ and $1.005627\times10^{-4}$, respectively.

\begin{figure*}[t]
    \centering
    \includegraphics[width=\textwidth]{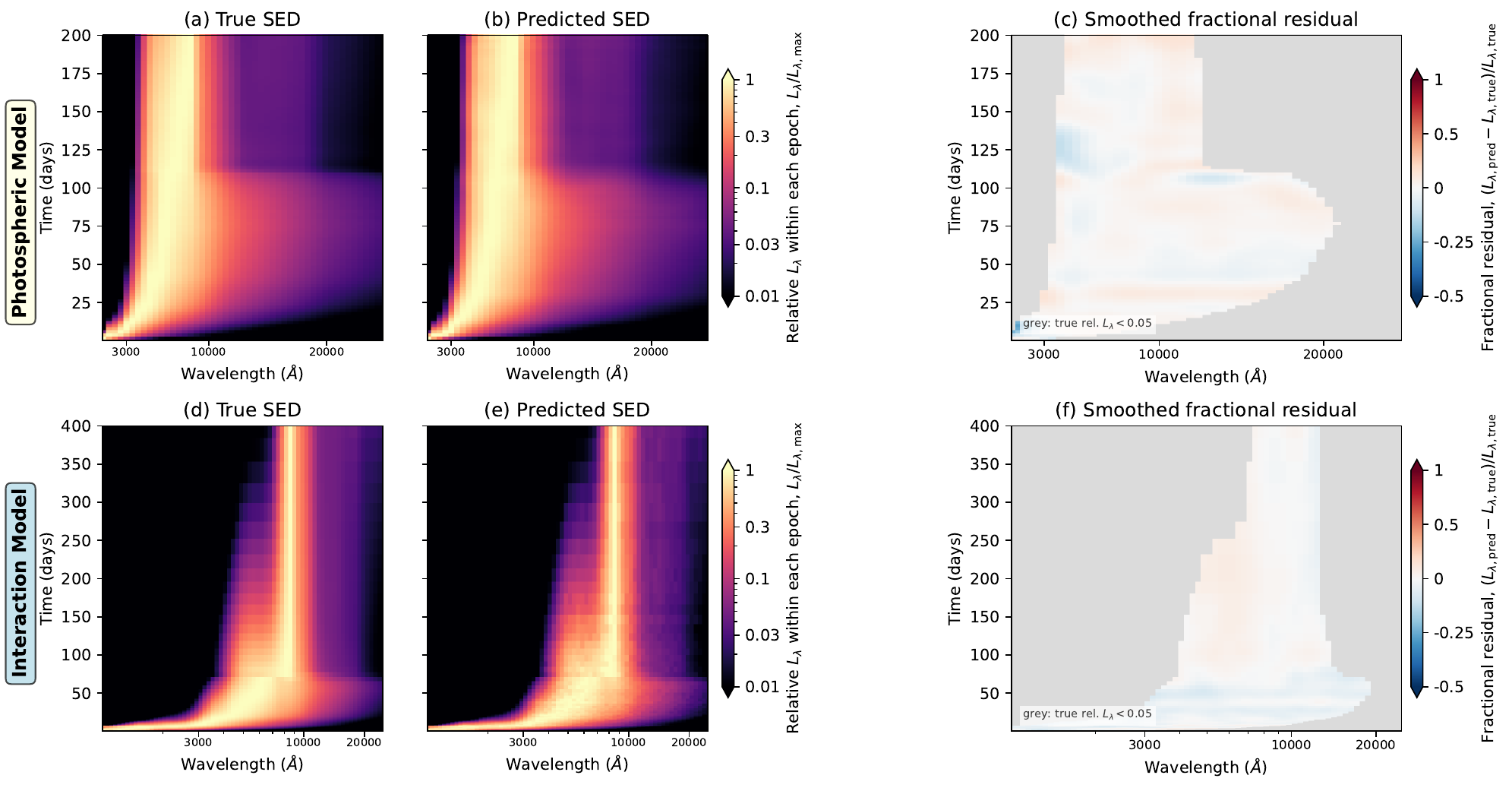}
    \caption{Spectral reconstruction performance on the held-out test sets. The top row (a--c) shows the photospheric model, while the bottom row (d--f) shows the interaction model. From left to right, the columns show the ground-truth \textsc{stella} SEDs, the emulator-predicted SEDs, and the residual maps. In the true and predicted SED panels, the color scale represents the dimensionless epoch-normalized relative luminosity, $L_\lambda(t,\lambda)/L_{\lambda,\max}(t)$, where $L_{\lambda,\max}(t)$ is the maximum $L_\lambda$ at the same epoch. Redder regions therefore correspond to wavelengths close to the normalized SED peak at that epoch. The residual panels show the fractional residual, $(L_{\lambda,\rm pred}-L_{\lambda,\rm true})/L_{\lambda,\rm true}$, computed from the per-epoch normalized relative $L_\lambda$ maps; low-signal regions with true normalized luminosity below 0.05 are shown in gray because fractional residuals in weak spectral tails are not informative. For visualization, the top-row photospheric model panels use a linear wavelength axis, matching the linear wavelength grid used for its training and preprocessing, whereas the bottom-row interaction model panels use a logarithmic (geometric) wavelength axis, matching the geometric wavelength grid used during training.
    \label{fig:spectra_comparison}}
\end{figure*}

Figure~\ref{fig:spectra_comparison} illustrates the reconstruction fidelity of our emulators across the full time--wavelength plane using the epoch-normalized quantity $L_\lambda/L_{\lambda,\max}$. This normalization is used only for visualization: it highlights the temporal evolution of the spectral shape. The progressive shift of the brightest normalized regions toward longer wavelengths is expected as type~II SN ejecta cool and the SED peak moves redward. The true and predicted two-dimensional SED maps agree closely for both the photospheric model (panels a--c) and the interaction model (panels d--f). The residual panels show the fractional residual $(L_{\lambda,\rm pred}-L_{\lambda,\rm true})/L_{\lambda,\rm true}$ in the same normalized space, with weak-tail pixels masked in gray where the true normalized luminosity is below 0.05.

\begin{figure*}[t]
    \centering
    \includegraphics[width=\textwidth]{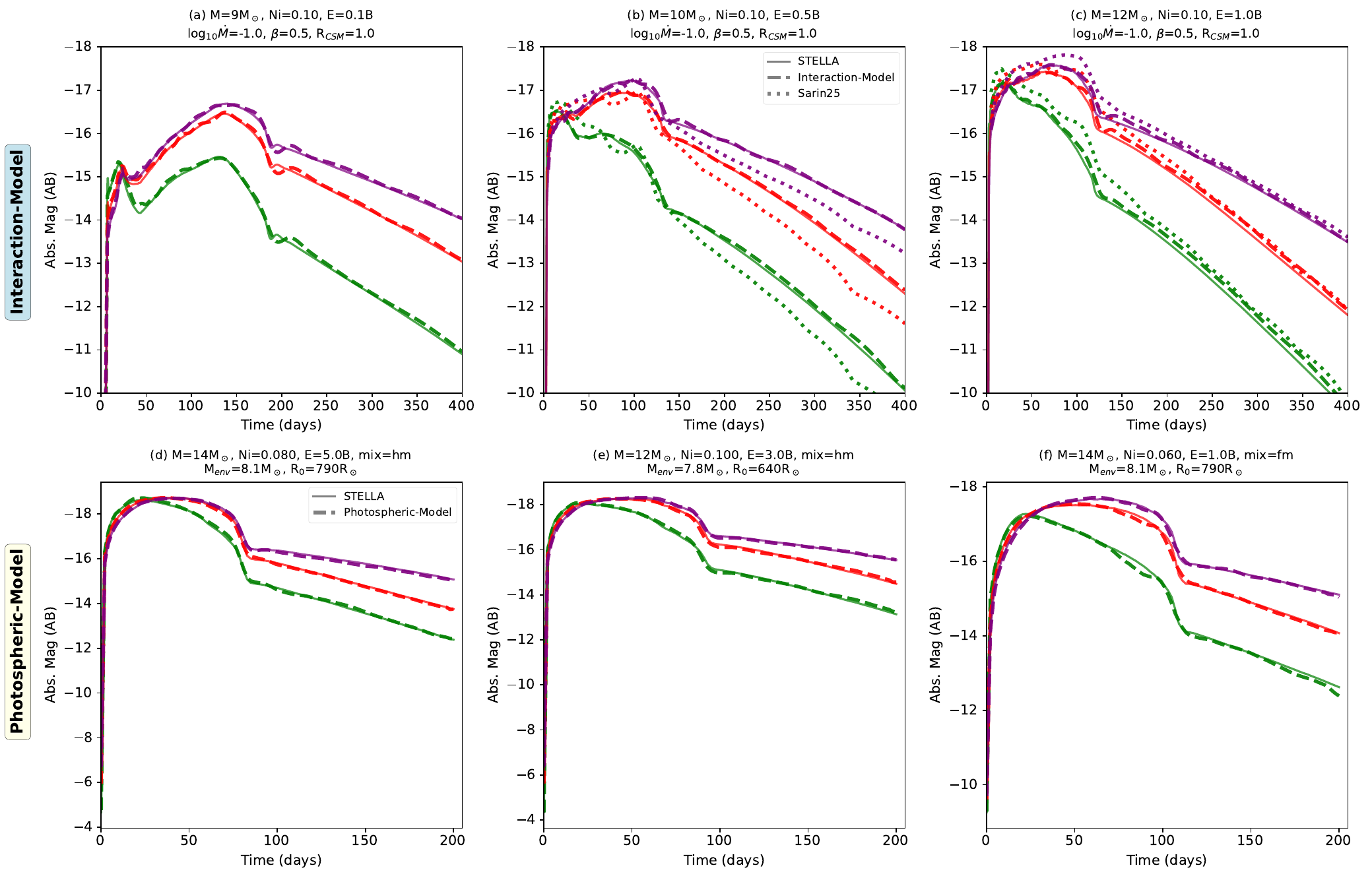}
    \caption{Comparison of multiband absolute magnitude light curves among ground-truth \textsc{stella} simulations (solid lines), our emulator predictions (dashed lines), and the Sarin25 model (dotted lines). The colors correspond to the ZTF filters: green for the $g$-band, red for the $r$-band, and purple for the $i$-band. Top row (interaction model): Validation in the low-mass, low-energy, CSM-interacting regime. Bottom row (photospheric model): Validation in the interaction-free regime.}
    \label{fig:lc_comparison}
\end{figure*}

This high fidelity at the spectral level translates directly into the accurate reconstruction of photometric observables, as illustrated in Figure~\ref{fig:lc_comparison}.

For the top-row interaction model cases, which sample the low-mass, low-energy, CSM-interacting regime, the predictions (dashed lines) closely track the ground-truth \textsc{stella} simulations (solid lines). As demonstrated in panels (a)--(c), the model maintains high fidelity across different parameter combinations, including both the rapid early rise and the transition phase.

For the photospheric model (bottom row), the emulator demonstrates a robust capability to capture interaction-free spectral features despite the limited training data, accurately reproducing the early cooling peak, the flat plateau phase, and the sharp transition to the radioactive decay tail.

\subsection{Parameter inference validation}

Before presenting the benchmark applications, we describe the inference setup used throughout this section. The benchmark fits are performed in broadband photometric space using the \textsc{redback}/\textsc{bilby} inference interface with the \texttt{dynesty} nested sampler. For a given parameter vector $\boldsymbol{\theta}$, the surrogate first generates a time-dependent \textsc{stella} SED. The SED is then convolved with the corresponding filter response functions to produce synthetic model magnitudes in the observed bands. These synthetic magnitudes are compared with the observed multiband photometry in the likelihood evaluation.

Both surrogate models are implemented through the \texttt{redback\_surrogates}\footnote{\url{https://github.com/nikhil-sarin/redback_surrogates}} interface to \textsc{redback}, an open-source Bayesian inference package for electromagnetic transients \citep{2024MNRAS.531.1203S}. The forward models support optional PyTorch GPU evaluation for generating synthetic light curves, but this acceleration acts at the forward-model level rather than as a unified native backend for the full \textsc{redback} sampling stack; consequently, the nested-sampling workflow used here is still primarily CPU executed.

The priors adopted in the benchmark inference are summarized in Table~\ref{tab:benchmark-priors}. Most physical parameters are assigned uniform priors over the range covered by the corresponding surrogate grid. The nuisance parameter $A$ is included to absorb residual uncertainty in distance, extinction, calibration, and model normalization.

\begin{table*}[t]
\centering
\caption{Priors used in the benchmark inference runs. `U'' denotes a uniform prior, `logU'' denotes a log-uniform prior, and ``Delta'' denotes a fixed value.}
\label{tab:benchmark-priors}
\begin{tabular}{l l l l}
\hline
Parameter & interaction model prior & photospheric model prior & Unit / note \\
\hline
$M_{\rm ZAMS}$ & U$(9,18)$ & U$(10,18)$ & $M_\odot$ \\
$E_{\rm SN}$ & U$(0.1,5)$ & U$(0.2,5)$ & $10^{51}$ erg \\
$M_{\rm Ni}$ & U$(0.001,0.3)$ & U$(0.0001,0.3)$ & $M_\odot$ \\
$\log_{10}\dot{M}$ & U$(-5,-1)$ & Not used & $M_\odot\, {\rm yr}^{-1}$ \\
$\beta$ & U$(0.5,5)$ & Not used & CSM density-slope parameter \\
$R_{\rm CSM}$ & U$(1,10)$ & Not used & $10^{14}$ cm \\
mixing & Not used & Categorical $\{0,1,2\}$ & $0={\rm cm}$, $1={\rm fm}$, $2={\rm hm}$ \\
$M_{\rm env}$ & Not used & U$(7.2,9.5)$ & $M_\odot$ \\
$R_0$ & Not used & U$(510,970)$ & $R_\odot$ \\
$A$ & logU$(0.01,2)$ & logU$(0.01,2)$ & Multiplicative flux scale \\
\hline
\end{tabular}
\end{table*}

We adopt a Gaussian likelihood in magnitude space. For each observed photometric data point $i$, the effective uncertainty is defined as
\begin{equation}
\sigma_{i,\rm eff}^{2} = \sigma_{i,\rm obs}^{2} + \sigma_{\rm add}^{2},
\end{equation}
where $\sigma_{i,\rm obs}$ is the reported photometric uncertainty and $\sigma_{\rm add}$ is a fixed additional uncertainty term added in quadrature to account for residual calibration, interpolation, and surrogate-model uncertainties. The log-likelihood is then
\begin{equation}
\ln \mathcal{L}(\boldsymbol{\theta}) =
-\frac{1}{2}\sum_{i=1}^{N_{\rm phot}}
\left[
\frac{\left(m_{i,\rm obs}-m_{i,\rm mod}(\boldsymbol{\theta})\right)^2}
{\sigma_{i,\rm eff}^{2}}
+
\ln\left(2\pi\sigma_{i,\rm eff}^{2}\right)
\right],
\end{equation}
where $m_{i,\rm obs}$ is the observed magnitude and $m_{i,\rm mod}(\boldsymbol{\theta})$ is the corresponding model magnitude generated from the surrogate SED after filter convolution. This likelihood definition is used consistently for all benchmark supernovae analyzed below.

We selected three benchmark supernovae and performed inference in broadband photometric space. Although our surrogates generate time-dependent \textsc{stella} SEDs, these SEDs are multigroup radiation-transfer outputs sampled with a finite number of frequency bins, and are therefore more appropriate for constructing synthetic broadband light curves than for direct line-profile-level comparison with observed spectra. For the interaction model, we analyzed SN~2005cs, which tests the robustness of the model in the presence of potential confined CSM interaction, as suggested by broad spectral ``ledge'' features. For the photospheric model, we employed the archetypal SN~1999em to characterize the standard regime dominated by pure ejecta evolution. Additionally, SN~2012aw was reanalyzed to facilitate comparison with previous model iterations.

\subsubsection{SN 2005cs}

\begin{figure*}[t]
    \centering
    \includegraphics[width=\textwidth]{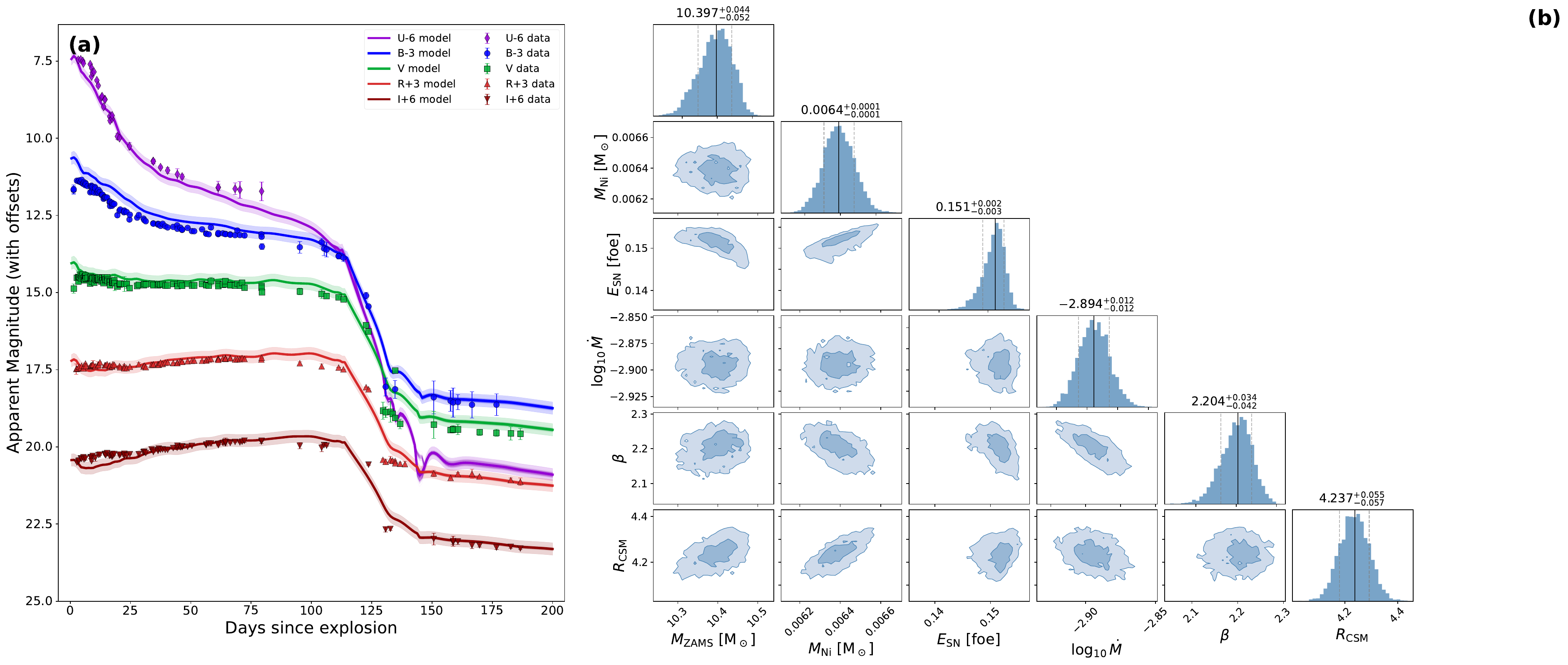}
    \caption{Parameter inference validation for SN 2005cs using the interaction model. 
    Panel (a): Comparison between the observed multiband photometry and the best-fit synthetic light curves (solid lines). The error bars represent a fixed 0.2 mag uncertainty around the light curves corresponding to the median inferred parameters.
    Panel (b): Corner plot displaying the posterior probability distributions for the inferred physical parameters.
    \label{fig:sn2005cs}}
\end{figure*}

SN~2005cs serves as a critical stress test for our interaction model because of the long-standing tension in its inferred physical parameters. The multiband photometric data used in our fit are adopted from the observations presented by \citet{2009MNRAS.394.2266P}. While direct analyses of preexplosion \textit{HST} images consistently support a low-mass progenitor ($6-10\,M_\odot$; \citealt{2005MNRAS.364L..33M,2006ApJ...641.1060L,2007MNRAS.376L..52E}), early hydrodynamic modeling required a significantly more massive envelope ($\sim 17.3\,M_\odot$) to reproduce the plateau properties \citep{2008A&A...491..507U}. Our inference results (Fig.~\ref{fig:sn2005cs}) offer a plausible resolution to this discrepancy. The interaction model provides a high-fidelity fit to the multiband light curves, tightly constraining the progenitor mass to $M_{\mathrm{ZAMS}} = 10.40^{+0.04}_{-0.05}\,M_\odot$ and the explosion energy to $E_{\mathrm{SN}} = 0.1515^{+0.0018}_{-0.0025} \times 10^{51}$\,erg. These values agree well with both direct imaging constraints and recent neutrino-driven explosion models \citep{2022MNRAS.514.4173K}, favoring the low-mass hypothesis. The inferred $^{56}$Ni mass, $M_{\mathrm{Ni}} = 0.00639^{+0.00009}_{-0.00008}\,M_\odot$, is also consistent with expectations for nickel-poor, low-energy iron core-collapse explosions.

Furthermore, the model characterizes the circumstellar environment as a confined dense shell ($R_{\rm CSM} = 4.237^{+0.055}_{-0.057} \times 10^{14}$\,cm) featuring an enhanced mass-loss rate ($\log_{10}\dot{M} = -2.894^{+0.012}_{-0.012}$) and a steep density gradient ($\beta = 2.204^{+0.034}_{-0.042}$). These properties provide a self-consistent physical framework for the interaction signatures observed in early spectra. This scenario receives critical observational support from the recent discovery of SN~2024abfl, a physical ``twin'' to SN~2005cs. Unlike the ambiguous early spectra of SN~2005cs—where the origin of similar features remained debated \citep{2008ApJ...675..644D}—SN~2024abfl clearly displayed a broad ``ledge'' feature near 4600\,\AA\ within days of explosion. This feature is interpreted as a blend of shock-accelerated high-ionization lines (e.g., He~II and N~III) arising from interaction with confined CSM \citep{2026arXiv260102638G,2026ApJ1002...68L}. This indicates that even low-mass red supergiant progenitors ($\sim 9$--$12\,M_{\odot}$) can experience enhanced wind mass loss. Consequently, SN~2024abfl provides compelling independent verification that the peculiar early spectral features of SN~2005cs were driven by CSM interaction rather than intrinsic ejecta properties.

\subsubsection{SN 2012aw}

\begin{figure*}[t]
    \centering
    \includegraphics[width=\textwidth]{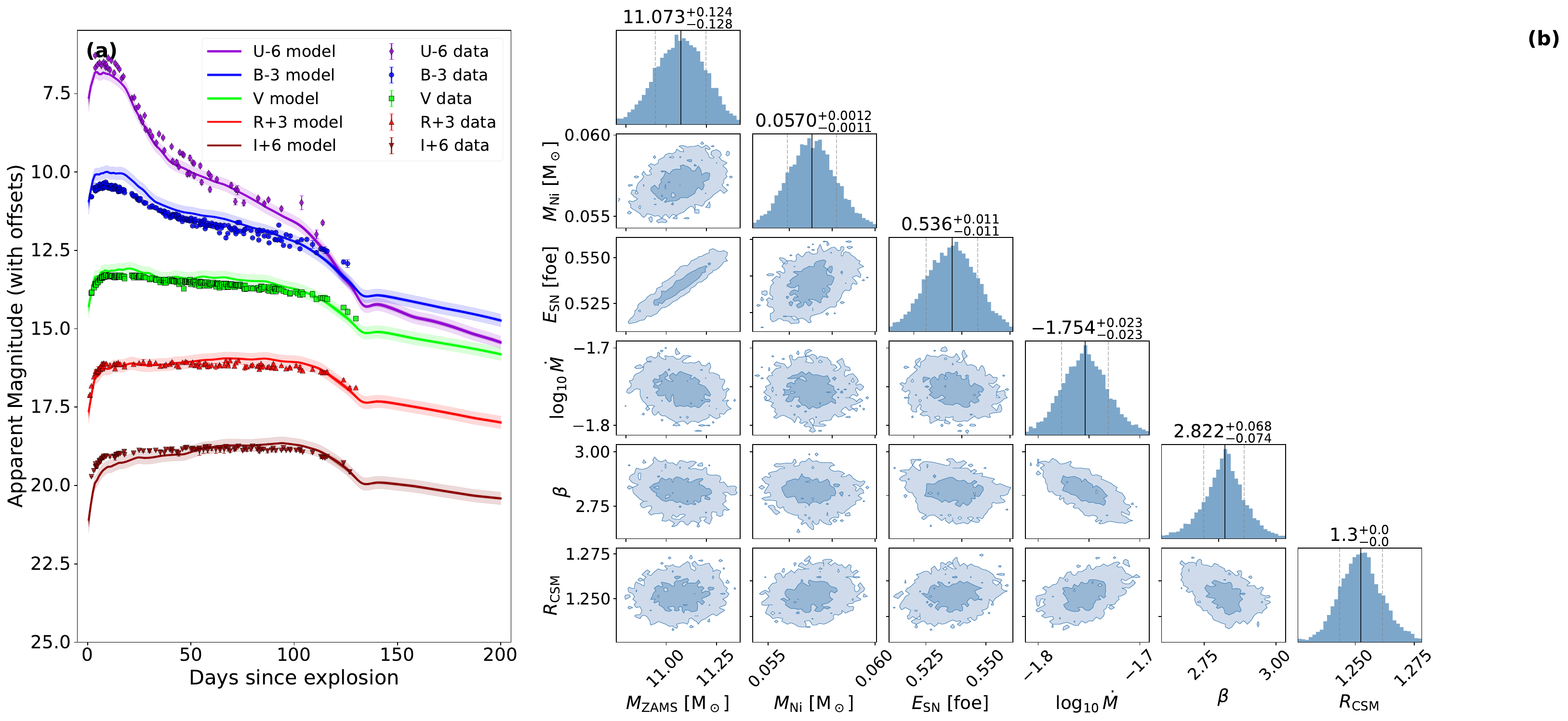}
    \caption{Same as Fig.~\ref{fig:sn2005cs}, but for SN 2012aw.
    \label{fig:sn2012aw}}
\end{figure*}

As illustrated in Figure~\ref{fig:sn2012aw}, we analyze SN~2012aw using the optical photometric dataset presented by \citet{2013MNRAS.433.1871B}. The inferred progenitor ZAMS mass ($M_{\mathrm{ZAMS}} = 11.05^{+0.06}_{-0.06}\,M_{\odot}$) agrees well with both direct progenitor-imaging constraints ($\approx 12.5\,M_{\odot}$; \citealt{2012ApJ...759L..13F,2012ApJ...759...20K,2016MNRAS.456L..16F}) and the estimate of \citet{2025MNRAS.544.2653S} ($10.61\,M_{\odot}$). For the $^{56}$Ni mass, our model yields a robust estimate: the derived $M_{\mathrm{Ni}} = 0.0557^{+0.0006}_{-0.0005}\,M_{\odot}$ closely matches the $0.06 \pm 0.01\,M_{\odot}$ measured by \citet{2013MNRAS.433.1871B} and is consistent with the $\sim 0.06\,M_{\odot}$ obtained by \citet{2014ApJ...787..139D}. We derive an explosion energy of $E_{\mathrm{SN}} = 0.533^{+0.005}_{-0.005} \times 10^{51}$\,erg, comparable to the $0.63_{-0.04}^{+0.05} \times 10^{51}$\,erg reported by \citet{2025MNRAS.544.2653S}. The high-velocity absorption features identified in early spectra by \citet{2013MNRAS.433.1871B} suggest ejecta-CSM interaction, while radio analysis by \citet{2014ApJ...782...30Y} reveals significant inverse Compton cooling, supporting a scenario in which the supernova resided in a dense environment with an intense radiation field. These findings support our inference that SN~2012aw is surrounded by a dense, confined CSM shell.

\subsubsection{SN 1999em}

\begin{figure*}[t]
    \centering
    \includegraphics[width=\textwidth]{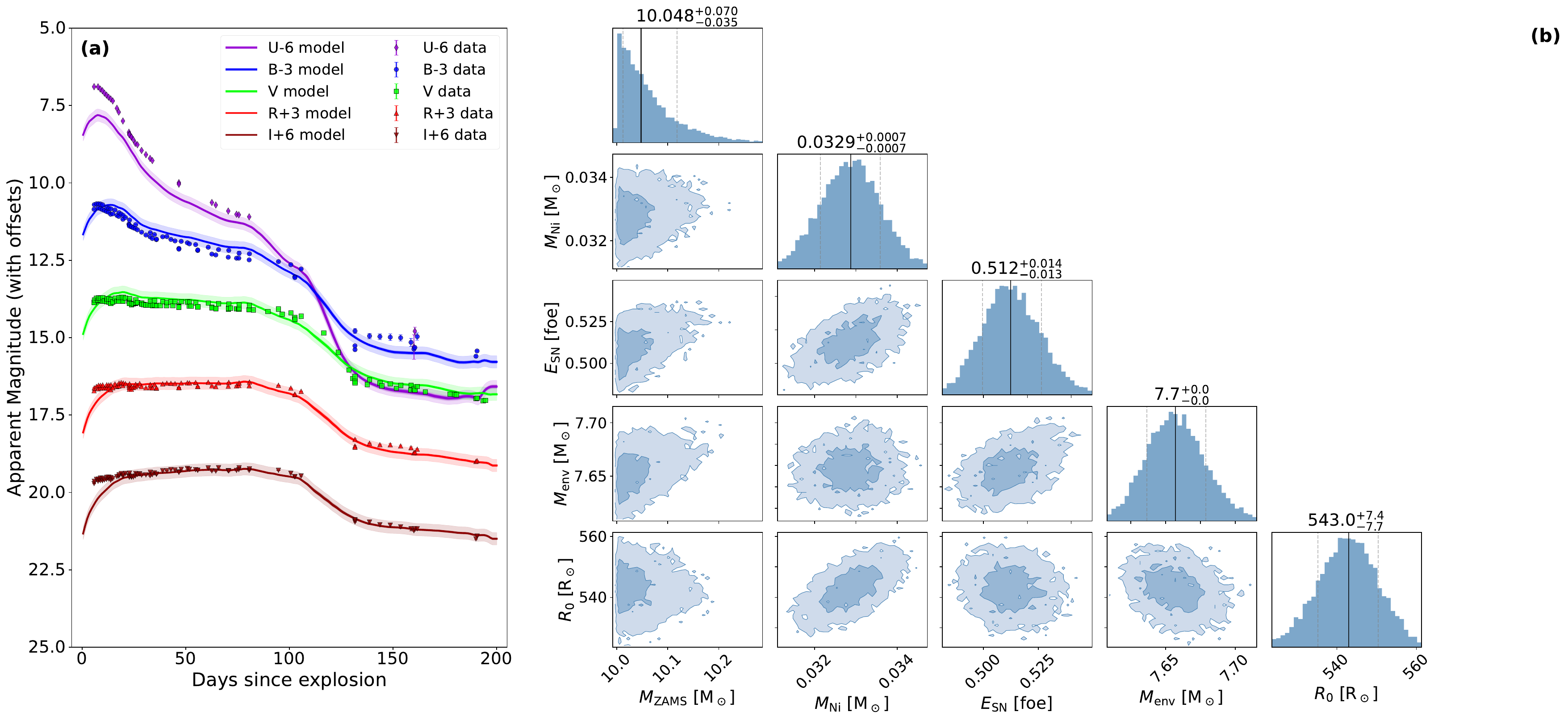}
    \caption{Same as Fig.~\ref{fig:sn2012aw}, but for the archetypal type IIP SN 1999em using the interaction-free photospheric model.
    \label{fig:sn1999em}}
\end{figure*}

SN 1999em is widely regarded as an archetype of normal type IIP supernovae, serving as an ideal benchmark for our interaction-free photospheric model. The photometric dataset used for this benchmark is taken from the optical monitoring presented by \citet{2003MNRAS.338..939E}. Our emulator reproduces the observed photometry reasonably without invoking explicit CSM interaction, deriving a progenitor mass of $M_{\mathrm{ZAMS}} = 10.05^{+0.07}_{-0.04} M_{\odot}$. This estimate is of particular interest regarding the historical mass discrepancy: while previous hydrodynamical studies (e.g., \citealt{2011ApJ...729...61B}) favored higher masses ($\sim 19 M_{\odot}$) to reproduce the light curve, our inference aligns with the direct preexplosion imaging limits ($12 \pm 1 M_{\odot}$; \citealt{2002ApJ...565.1089S}).

Regarding explosion properties, we derive a synthesized $^{56}$Ni mass of $M_{\mathrm{Ni}} = 0.033^{+0.001}_{-0.001} M_{\odot}$ with negligible mixing. Our derived values are broadly consistent with previous photometric and spectroscopic estimates in the literature \citep{2003MNRAS.338..939E,2006A&A...447..691D}. This photospheric model interpretation is also consistent with the literature view of SN\,1999em as a normal type~IIP event: its photometric and spectroscopic evolution was extensively studied by \citet{2003MNRAS.338..939E}, and its bolometric light curve and H$\alpha$ evolution can be reproduced by conventional hydrodynamic/photospheric modeling \citep{2007A&A...461..233U}. In addition, x-ray and radio observations indicate only weak ejecta--CSM interaction with a low-density red-supergiant wind, corresponding to a mass-loss rate of order $10^{-6}$--$10^{-6.5}\,M_\odot\,{\rm yr}^{-1}$ rather than a dense compact CSM component dominating the optical light curve \citep{2002ApJ...572..932P,2007ApJ...662.1136C}. These results demonstrate that the photospheric model captures the essential physics of standard type IIP explosions. However, we also observe a systematic discrepancy during the first $\lesssim 10$ days, where the observed light curve is $\approx 0.3 - 0.6$ mag brighter than our best-fit interaction-free model. This excess may reflect weak CSM interaction or other effects not included in the present photospheric framework, such as cooling-envelope emission, uncertainties in the early color/temperature evolution, line-blanketing effects, or limitations of the simplified model assumptions. Nevertheless, the photospheric model remains a viable alternative for characterizing type IIP events, especially when focusing on the main plateau phase.

\section{Discussion and conclusion}\label{sec4}

While our emulators demonstrate high statistical precision, their physical accuracy is fundamentally bounded by the underlying training data generated by \textsc{stella}. A primary limitation is the assumption of local thermodynamic equilibrium (LTE). This approximation holds well during the optically thick photospheric phase but breaks down as the ejecta expand and transition into the nebular phase. As explicitly noted by \citet{2025MNRAS.544.2653S}, LTE-based predictions become increasingly unreliable at epochs $\gtrsim 100$ days postexplosion. Consequently, any inference derived from the late-time radioactive decay tail should be interpreted with caution. Future iterations could integrate non-LTE codes like \textsc{cmfgen} or \textsc{sumo} to address this regime, although the computational cost of generating dense non-LTE grids remains a significant challenge.

The applicability of the interaction model is also bounded by the parameter space of the training grid, specifically the upper mass-loss rate limit of $\dot{M} = 10^{-1.0}\,M_{\odot}\,\mathrm{yr}^{-1}$. This range is sufficient for events with weak interaction, but it is not suitable for characterizing extreme interaction events such as SN\,2013fs \citep{2017ApJ...838...28M}. Such objects typically require significantly higher CSM densities ($\dot{M} > 0.15\,M_{\odot}\,\mathrm{yr}^{-1}$) to reproduce their strong, persistent narrow emission lines---a regime that remains to be explored in future grid expansions.

In summary, to address the analysis bottlenecks for massive type II supernova datasets, we have presented two independent neural network surrogate models---the interaction model and the photospheric model---designed for distinct physical regimes. The interaction model targets low-energy explosions with possible CSM interaction, whereas the photospheric model provides dedicated support for standard interaction-free events. We tailored specific machine learning architectures to these regimes: the interaction model utilizes a ResNet-based autoencoder for complex interaction scenarios, while the photospheric model employs a 2D CNN optimized for SED reconstruction of standard events. Crucially, both architectures incorporate ``latent mixup'' regularization to effectively smooth the high-dimensional latent space. We summarize the posterior medians and 68\% credible intervals for the three benchmark supernovae in Table~\ref{tab:sn_params}, which demonstrates that the surrogates recover physically plausible progenitor, explosion, and CSM properties across both interaction and interaction-free regimes.

\begin{table*}
\caption{Posterior medians and 68\% credible intervals for the physical parameters derived from our surrogate models.\label{tab:sn_params}}
\centering
\begin{tabular}{lcccccc}
\hline\hline
\multicolumn{7}{c}{Interaction model} \\
\hline
\multicolumn{1}{c}{Name} & $M_{\rm ZAMS}$ [$M_\odot$] & $M_{\rm Ni}$ [$M_\odot$] & $E_{\rm SN}$ [foe] & $\log_{10}\dot{M}$ & $\beta$ & $R_{\rm CSM}$ [$10^{14}$ cm]\\
\hline
SN\,2012aw & $11.05^{+0.06}_{-0.06}$ & $0.0557^{+0.0006}_{-0.0005}$ & $0.533^{+0.005}_{-0.005}$ & $-1.76^{+0.010}_{-0.010}$ & $2.82^{+0.02}_{-0.03}$ & $1.250^{+0.004}_{-0.004}$\\
SN\,2005cs & $10.40^{+0.04}_{-0.05}$ & $0.00639^{+0.00009}_{-0.00008}$ & $0.1515^{+0.0018}_{-0.0025}$ & $-2.894^{+0.012}_{-0.012}$ & $2.204^{+0.034}_{-0.042}$ & $4.237^{+0.055}_{-0.057}$\\
\hline
\multicolumn{7}{c}{photospheric model} \\
\hline
\multicolumn{1}{c}{} & $M_{\rm ZAMS}$ [$M_\odot$] & $M_{\rm Ni}$ [$M_\odot$] & $E_{\rm SN}$ [foe] & $M_{\rm env}$ [$M_\odot$] & $R_{0}$ [$R_\odot$] & mixing\\
\hline
SN\,1999em & $10.05^{+0.07}_{-0.04}$ & $0.033^{+0.001}_{-0.001}$ & $0.512^{+0.014}_{-0.013}$ & $7.66^{+0.02}_{-0.02}$ & $543^{+7}_{-8}$ & $0.0$\\
\hline\hline
\end{tabular}
\end{table*}

These surrogates are distributed through \texttt{redback\_surrogates}. In this implementation, a full Bayesian fit for a single type~II supernova can be completed in minutes rather than the days typically required when each likelihood evaluation calls a radiation-hydrodynamics simulation. We are currently applying these surrogates to well-observed type II supernova samples from the CSP and ZTF surveys \citep{ZhangInPrep}. This statistical application paves the way for the real-time physical characterization of the tens of thousands of SNe II expected annually from LSST \citep{2019ApJ...873..111I}, thereby enhancing our capability to derive statistical constraints on massive star explosions.

\begin{acknowledgments}
We thank the anonymous reviewer for the valuable comments and suggestions. 
We acknowledge the Stellar Physics Group of Yunnan Observatories for providing computational resources. This study was supported by the National Natural Science Foundation of China (Nos. 12288102, 12225304, 12090040/12090043), the CAS Project for Young Scientists in Basic Research (No. YSBR-148), the National Key R\&D Program of China (No. 2021YFA1600404), the science research grant from the China Manned Space Project (No. CMS-CSST-2021-A12), the Yunnan Revitalization Talent Support Program (Yunling Scholar Project), the Yunnan Fundamental Research Project (Nos. 202201BC070003, 202501AS070005), the Yunnan Science and Technology Program (Nos. 202605AS350010 and 202601BC070011), and the International Centre of Supernovae (ICESUN), Yunnan Key Laboratory of Supernova Research (Nos. 202302AN360001 and 202505AV340004). S.Z.\ is supported by the National Natural Science Foundation of China (Nos. 12393811, 12473031), the Yunnan Revitalization Talent Support Program (Young Talent Project), and the Yunnan Fundamental Research Project (No. 202501AS070078). C.\ Wu is supported by the National Natural Science Foundation of China (No. 12473032), the Yunnan Revitalization Talent Support Program (Young Talent Project), and the Yunnan Fundamental Research Project (Nos. 202501AW070001, 202301AU070039).
\end{acknowledgments}

\section*{Data availability}
The software used in this work is publicly available through \texttt{redback\_surrogates} at Ref.~\cite{redbackSurrogates}. Requests for other data or further information should be sent to the authors.

\appendix

\section{Validation of latent space smoothness}
\label{app:latent_smoothness}

For the emulator to accurately map continuous physical parameters to spectra, the underlying latent space (Stage 1) must be smooth and free of discontinuities. A fragmented latent space would lead to prediction artifacts.

\begin{figure*}
    \centering
    \includegraphics[width=\textwidth]{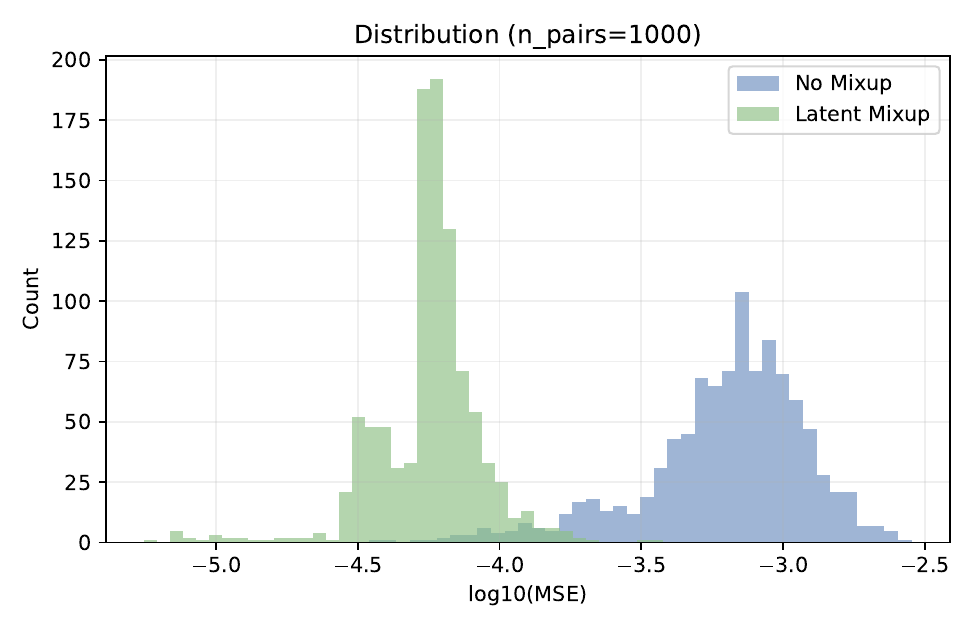}
    \caption{Latent space interpolation test. We compare the reconstruction error of latent midpoints for 1,000 random pairs. The latent mixup model (green) exhibits an order-of-magnitude reduction in MSE compared to the baseline (blue), indicating a significantly smoother and more linear latent manifold.
    \label{fig:latent_interp}}
\end{figure*}

To verify the smoothness of our learned manifold, we performed a ``Midpoint Interpolation Test.'' We randomly sampled $N=1,000$ distinct pairs from the test set. For each pair $(x_1, x_2)$, we compared the \textit{decoded latent midpoint} against the \textit{linear average} of the inputs using the MSE:

\begin{equation}
    \mathcal{L}_{\text{interp}} = \left\| \mathcal{D}\left(\frac{\mathcal{E}(x_1) + \mathcal{E}(x_2)}{2}\right) - \frac{x_1 + x_2}{2} \right\|^2
    \label{eq:interp_loss}
\end{equation}

Figure \ref{fig:latent_interp} shows the distribution of this error. The model trained with latent mixup (green) achieves an interpolation error approximately one order of magnitude lower than the baseline (blue), with the median MSE dropping from $\sim 6.0 \times 10^{-4}$ to $\sim 5.0 \times 10^{-5}$. This confirms that latent mixup effectively enforces a continuous and well-interpolatable latent structure, ensuring robust spectral generation.

\clearpage

%

\end{document}